\documentclass[letterpaper]{article}

\usepackage{natbib}
\usepackage{microtype}
\usepackage[hidelinks]{hyperref}
\usepackage{caption}
\usepackage{todonotes}
\usepackage{url}
\usepackage{tikz}
\usepackage{amsmath}
\usepackage{xcolor}
\usepackage{graphicx}
\usepackage{enumitem}
\usepackage{csquotes}
\newcommand\crule[3][black]{\textcolor{#1}{\rule{#2}{#3}}}
\usetikzlibrary{positioning}
\usetikzlibrary{arrows}
\usetikzlibrary{automata}
\usepackage{pgfplots}
\usetikzlibrary{positioning}

\title{  Time as It Could Be Measured in Artificial Living Systems}
\author{Andrei D. Robu$^{1}$, Christoph Salge$^{1,2}$, Chrystopher L. Nehaniv$^{1}$ \and Daniel Polani$^{1}$ \\
\mbox{}\\
$^1$Adaptive Systems Research Group, School of Computer Science\\
University of Hertfordshire, Hatfield, UK  \\
$^2$Game Innovation Lab, Department of Computer Science and Engineering\\
New York University, New York, USA \\
a.robu@herts.ac.uk}

\begin{document}

\newcommand{\ask}[1]{\todo[inline,color=teal]{#1}{}}
\newcommand{\lang}[1]{\todo[inline,color=pink]{#1}{}}
\newcommand{\fix}[1]{\todo[inline]{#1}{}}

\newcommand{\diffnew}[1]{\ifdefined\DIFadd\DIFadd{#1}\else#1\fi}

\maketitle

\begin{abstract}
Being able to measure time, whether directly or indirectly, is a significant advantage for an organism. It permits it to predict regular events, and prepare for them on time. Thus, clocks are ubiquitous in biology.
In the present paper, we consider the most minimal  abstract pure  clocks  and investigate their characteristics with respect to their ability to measure time. Amongst other, we find fundamentally diametral clock characteristics, such as oscillatory behaviour for local time measurement or decay-based clocks measuring  time periods in scales global to the problem. We include also cascades of independent clocks (``clock bags'') and  composite clocks with controlled dependency; the latter show various regimes  of markedly different dynamics.
\end{abstract}

\section{Introduction}
``Bacteria count on probabilistic fingers'', to put it crisply, and count they must when they measure time. In the present paper, we study how such counting/time measurement could possibly look like in the most minimal cases which we believe relevant to simple organisms. \cite{klyubin2007representations} look at maximising information flow for a navigation task. They find that it is possible to obtain information about the current time. While there time measurement was a side effect, here we make it our primary objective: measure time as the most intrinsic, least environmentally affected processual quantity, with only three assumptions: it is discrete, the tick is global (accessible to the agent) and time has a well-defined beginning.
More precisely, we are interested in how much a Markovian agent with limited memory can keep track of the flow of time under these assumptions. It turns out that, in this context, measuring time is effectively counting. To make the best out of limited resources, we need to count probabilistically.

In its conceptually simplest incarnation, the measurement of time would consist essentially of two components: Having a reliable generator of periodic behaviour on the one hand; and being able to count the periods, on the other. To measure larger time intervals precisely, one needs full-fledged counters, which, in turn, require a comparatively complex logical make-up; something which, while not impossible in principle, one would not expect generically in biologically relevant scenarios and most certainly not in very simple organisms. Rather, one would typically expect to find some less precise, but simpler and more robust solution. The present paper investigates how the most minimal of such models could look and which characteristic properties we expect them to exhibit.
In our discussion we include both explicit clocks characterized by a  distinct apparatus with a clear function such as the suprachiasmatic nucleus in the mammalian hypothalamus involved in the control of circadian rhythms \citep{klein1991suprachiasmatic}, but also implicit clocks (which exhibit some aspects of clock-like behaviour without a dedicated mechanism) like the feeding-hunger cycle of an animal. We do not differentiate these classes a~priori since our formalism does not discriminate between them, but our study will concentrate on systems which, given their constraints, are maximally able to measure time, not considering any other tasks --- thus we study clocks as ``pure'' as they can be under the circumstances given. We expressly are not studying how an agent could infer time from correlations in the environment (using a sundial for example), but only in how time can be tracked intrinsically. In the same vein as considering Artificial Life as about understanding life-as-we-know-it vs.\ life-as-it-could-be \citep{langton89:_artif_life}, and in view of how life pervasively makes use of clocks, the present paper studies possible clocks themselves, or, as we say in the title: ``time-as-it-could-be-measured''.

\section{Temporal Dynamics: The Algebra of Time}

We begin with some general comments on the structure of time. Whether in classical or relativistic physics, or
in more general models (e.g., ancient Indian notions of cyclical time,
or in modern automata networks), there is a commonality to notions and
models of time: 
From the perspective of a single organism or agent,
events in time satisfy a grammatical constraint \citep{rhodes2009applications}:
If $\alpha$, $\beta$ and $\gamma$ are each sequences of events, then:
if $\beta$ follows $\gamma$ in an agent's experience, and, prior to both, $\alpha$ occurs,
that is exactly equivalent to when $\beta$ follows $\alpha$, and $\gamma$ occurs after both, i.e.\ $
\alpha(\beta \gamma) =  (\alpha \beta) \gamma$.

In short,  sequences of events in time from the perspective of an  individual agent
satisfy the  {\em associative law}.  Thus, the study of (possibly general) structures satisfying
this law becomes the study of models of time.      This viewpoint has deep connections
with the theory of discrete or continuous dynamical systems \citep{rhodes2009applications}. In the finite discrete deterministic case it leads to the Krohn-Rhodes theory,
a branch of mathematics (algebraic automata theory) where discrete dynamical
systems can be decomposed using (non-unique) iterative coarse-graining into a cascade of
irreducible components. The composite clocks discussed later constitute a special case of such a decomposition.

The simplest models of time have a single discrete kind of event,  a {\em clock tick}, which drives
their dynamics deterministically. The dynamics can be classified into four types \citep{nehaniv1993algebra}:  
\emph{cycles} (after  a number $n$ of steps everything repeats), \emph{fuses} (after a number of steps $k$ nothing ever
changes again), 
fuses that end in cycles (fuses that after $k$ transient clock ticks end in a cycle, generalizing the former two cases), or infinite systems with time indexed by the natural numbers
(where each moment is different from the others).  These types correspond to four different kinds of algebraic structures called \emph{semigroups} with a single generating event. 
If more than one type of event is allowed (e.g., clock ticks of different types, or sensory input with more than 1-bit), or if the state transitions are not deterministic, much more complicated and interesting dynamics can arise.
Here we seek to understand the structure of single tick time ``as it could be'' for agents experiencing it in a probabilistic or noisy setting, using the language of
 information theory for selected basic cases. Thus the rest of the paper expands beyond deterministic models into probabilistic ones.
  
\section{The Cost of Measuring Time}
Recent work by  \citet{barato16:_cost_precis_brown_clock}  highlighted interest in the problem of time measurement by asking if clocks must pay a thermodynamic cost to run. Their stance is grounded in fundamental trade-offs of physics. Our perspective at the level of organisms is far remote from these trade-offs, and  physical intuitions and conservation laws and constraints do not apply in a straightforward manner.
Instead, we work in a near-macroscopic, classical (non-quantum) Markovian universe without presuming the additional structure of microphysics (microreversibility or Hamiltonian dynamics). In particular, we cannot assume an obvious generalization of the physical concept of energy to a fully general Markovian system. Thus, it is not obvious how to cost computation in terms completely analogous to thermodynamics; the only concepts that carry over are of entropic/information-theoretic nature. Shannon information has been shown to be a highly generalizable measure of information processing cost \citep{polani2009information}; a universal measure that can even be used to compare systems of a different nature, it does not presuppose any structure on the state space of events.
All costs will, therefore, be expressed in this language, specifically information storage and communication costs. 

\emph{Small State Spaces}. We begin by focusing on minimal clocks, for a number of reasons. For one, this will make it easier to explore the full solution space. The second reason is significantly more subtle, and we will only be able to sketch it here: essentially, it is not clear what measure of complexity to utilize for the cost of running a larger single-component counter and/or the complexity of running the transition itself; natural candidates for such costs might be predictive information \citep{bialek01:_predic} or statistical complexity \citep{crutchfield89:_infer}. However, the possible candidates are not limited to these two measures, and many other plausible information-theoretic alternatives can be conceived\footnote{We will investigate this question in a separate paper.}. In absence of a canonical measure for the complexity of a single-component clock, here we limit ourselves to investigate the most minimal  clocks possible, namely a 2-state (1-bit) clock  and we will here not further concern ourselves to take into account the informational cost of actually running this clock.

\emph{Information flow between modules}. We said above that we prefer minimal clocks by default, to avoid dealing with the complexity of the clock operation. But what if an agent needs a larger clock? In this case, we build it out of smaller clocks. Such  a compound clock will perform better if its components ``cooperate''. For them to be able to cooperate, they must be able to exchange information. Therefore limitations in this clock's internal information flow will reduce the performance of the clock. Yet this information flow will in general be costly \citep{laughlin01:_energ}. This will limit how many components can communicate with each other and at which bandwidth. All in all, in our studies, when looking for candidates for clocks, we prefer Markov chains with a small number of states and when we move on to larger clocks we prefer to build them out of small clocks and use the information flow between the components as cost. \section{Other Relevant Work}
\cite{klyubin2007representations} consider the maximization of information flows in a very simple agent/environment system. The resulting agent controllers generate a rich set of behaviours. One of the side effects of the controllers' dynamics is that their internal states partially encode location information, but also partially time information. In other words,  the resulting agent controllers provide partial information about the point in time in the experiment. Amongst other, the paper studied how space and time are encoded together and how and to which extent they can be separated (``factorized'') in informational terms. 
Note that in \citep{klyubin2007representations}, the measurement of time was a side effect of the overall information flow optimization. The importance of measuring specifically time (as opposed to having this measurement emerge on the side) appears in other scenarios, for instance in 13- or 17-year cicadas \citep{karban00:_how,sota13:_indep}.  On the level of physical limits, \citet{chen2010clocks} study the statements that can be obtained by applying Fisher information to the problem of measuring time with quantum clocks and discuss the problem of clock synchronization in the quantum realm.

\cite{karmarkar2007timing} note mounting evidence against the view that the brain keeps track of time by counting ticks and instead study the behaviour of simulated SDNs (state-dependent neural networks), showing how neural networks can keep track of time implicitly in their states without counting. \Citet{van2012dynamic} discusses not only the possible mechanisms of time perception in the brain but also the difficulty of validating them.

Since we hypothesize that our abstract considerations find functional correspondences in nature, the simplest examples would be expected to be found in microorganisms. We would predict that, if our general hypothesis is correct, the characteristics of our results for minimal clocks will be reflected in very simple organisms. We will preempt here one result detailed in the results section. Under constraints on the memory available to the clock, only two types of clocks are found --- local, short-term clocks  measuring the time within a cycle (essentially its phase); and, long-term clocks which distinguish large-scale phases within the overall time interval of interest. Thus, we get a dichotomy between local time measurement and global time measurement. We will call the first type cyclic clocks or oscillators, and the second type ``drop clocks'' (essentially one-off decay-type time measurements). Bacteria show examples of both cyclic and drop clocks.

\citet{hut2011evolution} discuss the reasons why day and night require different behaviours, these forming the selective pressures of evolution for the circadian rhythm of bacteria --- the periodic cycle of activity in organisms allows them to adapt to the time of day. This is one example demonstrating the importance of time-keeping to organisms. They note that the circadian rhythm of cyanobacteria has been reproduced and studied in a test tube. Therefore this is an example of a relatively simple, clearly understood explicit cyclic or \emph{alternator} (see below) clock. 
\citet{nutsch2003signal} study prokaryotic taxis for halobacteria, more specifically signal transduction pathway starting from sensing light and responsible for controlling the switching of the flagellum. This pathway implements a one-off drop clock\footnote{Which is reset only by an external trigger; some mechanisms can be considered drop clocks  triggered at conception and never reset until death, such as telomeres.}.  They note that while the molecules involved in this pathway are well studied, the way these components behave together dynamically is not well understood and only speculated on. Building upon an earlier model by \cite{marwan1987signal}, they suggest dynamical models to fit experimental findings. These models are synthetic and not based on first principles. Thus, the question about the actual structure and size of the fundamental clocks of  bacteria remains currently unanswered. 

Measuring time can also help bacteria in spatial tasks.  Some bacteria use differences along the length of their body to measure gradients \citep{oliveira2016single}, however, others instead  measure time differences between intensities while moving \citep{nutsch2003signal}. This is demonstrating how measurement of space can be converted into measurement of time. 
\section{Models of Clocks}
\emph{\textelp{} time is always encoded in physical systems; \textelp{} any evolving system with nontrivial dynamics can be regarded as a clock. To read the time, we must perform measurements \textelp{}. How precisely we can estimate \textins{time} characterizes the quality of a clock.} ---~\cite{chen2010clocks}

Because our clocks are Markov chains, their memory is explicitly encoded in their current state and they are probabilistic. We consider only discrete state space and also advance in time only in discrete steps; importantly, this means that they receive time ticks for free and that they need not be concerned with the challenge of getting accurate ticks (see also discussion). Apart from these global ticks, the clocks do not receive any information from the environment. We furthermore consider first only the most minimal clock designs. We will later consider more complex (composite) clocks.

First, introduce some notation. Random variables are written as capital, their values as lowercase letters. The probability of a random variable $X$ to adopt value $x$ will be written as $P(X=x)$ or, where not ambiguous, as $p(x)$ by abuse of notation.
The state of a clock will be denoted by random variables $S$ (possibly subscripted by time $t$, because the probability distribution over the states of the clock changes in time) which can take on values $u$ and $d$ (``up'' and ``down''). To model the uncertainty that the agent has about the current time, we treat true time (which the clocks attempt to measure), similarly to \cite{klyubin2007representations} also as random variable $T$ which a~priori assumes all possible (integer) time values between $t=0$ and $t=T_{\text{max}}$ with equal probability.

Typical quantities would now be, e.g.\ 
$P(S=u|T=t) \equiv p(u|t)$, the probability that the clock state is $u$ (``up'') at time $t$. Or else, $P(T=t|S=d) \equiv p(t|d)$ would be the probability that the time is $t$, given that the current state of $S$ is $d$, etc. When we optimize clocks, we quantify our criterium of "performance" as mutual information $I(S;T)$. This mutual information tells how much information a previously uninformed agent would receive about time after looking at the clock (averaged for the possible observations of the clock, ``up'' or ``down''). This quantity, $I(S;T)$ is directly related to the probability of guessing time correctly given one observation of the clock. With this notation, we are ready to define the clocks. In all our experiments, the clocks will be initialized at time $t=0$ in a fixed known state, specifically $u$, i.e.\ $P(S_0 = u) \equiv P(S=u|T=0) = 1$ and advance once in time per tick according to their respective dynamics.

\subsection{Alternator}
\label{alternator-clocks}
We first define the \emph{alternator} as the 2-state Markov chain with a symmetric probability to change state.

\begin{figure}[htbp]
\centering
\begin{tikzpicture}[>=stealth',shorten >=1pt,auto,node distance=2cm]
  \node[initial,state] (u)      {$u$};
  \node[state]         (d) [right of=u]  {$d$};

  \path[->]
      (u) edge [loop above] node {$1 - r$} (u)
      (u) edge [bend left]  node {$r$} (d)
      (d) edge [loop above] node {$1 - r$} (d)
      (d) edge [bend left]  node {$r$} (u);
\end{tikzpicture}
\caption{The alternator clock. It switches one state to the other with probability $r$.}
\end{figure}
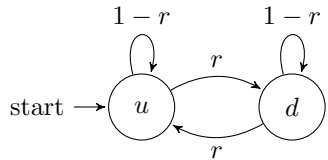

\begin{figure}[htbp]
\centering
\begin{tikzpicture}
\begin{axis}[
    ymin = 0, ymax = 1,
    xmin = 0, xmax = 20,
    xlabel = $t$,
    ylabel = {$p(u \mid t)$},
]
\addplot [
    domain=0:20, 
    samples=21
]
{0.5*(1+(1-2*0.95)^x)};
\addlegendentry{$r = 0.95$}
\addplot [
    domain=0:20, 
    samples=21, 
    dashdotted
]
{0.5*(1+(1-2*0.05)^x)};
\addlegendentry{$r = 0.05$}
\addplot [
    domain=0:20, 
    samples=21, 
    dashed
]
{0.5*(1+(1-2*0.5)^x)};
\addlegendentry{$r = 0.5$}
\end{axis}
\end{tikzpicture}
\caption{The alternator clock. Behaviour classes explained in text.}
\label{fig:alternator-time-behaviour}
\end{figure}
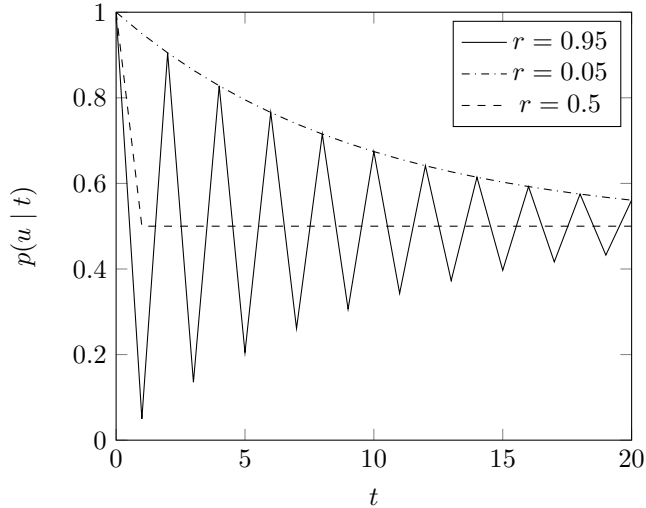

The plot of ${p(u \mid t)}$ in
Figure~\ref{fig:alternator-time-behaviour}, shows the 
distinction between different regimes in which
the clock operates, depending on $r$. We notice the analogy with
Damped Harmonic Oscillation: \emph{Stuck}, for ${ r = 0 }$, the clock is stuck in state $u$ (not shown in diagram). \emph{Overdamped}, for ${ 0 < r < 0.5}$, the probability distribution behaves like an overdamped oscillator. It is interesting to note that, in this regime, the probability distribution of the alternator clock has the same envelope as the drop clock which we discuss below.\emph{Critically Damped}, for ${ r = 0.5 }$, the clock acts analogous to a critically damped oscillator: it reaches the  equilibrium in the shortest possible amount of time --- one time step. \emph{Underdamped}, for ${ 0.5 < r < 1 }$, the probability distribution behaves like an underdamped oscillator.
This can be interpreted as the clock being initially ``synchronized'' with time like a clock, but that
the synchronization gets lost as the clock evolves, until the correlation between $t$ and $s$ disappears. \emph{Undamped}, for ${ r = 1 }$, the clock state alternates between $u$ and $d$. Not shown in figure. These behaviours are taken from the plot of the symmetric alternator, but the asymmetric one (with different transition probabilities $u \rightarrow d$ and $u \leftarrow d$) has the same qualitative behaviour, just with a different equilibrium point. All 2-state Markov chains belong to one these 5 classes. We will refer to this insight in the results section again.

\subsection{Drop Clock}
\label{drop-clock}
Consider the thought experiment of an insect that leaves its nest to forage or explore. Even if it does not find anything, the insect should still return at one point, otherwise it risks getting lost. This insect would profit from  a clock  telling it if ``it's been awhile''. An alternator is not well suited for this task; it  measures a local phase (an odd or even step), but it does not  provide much ``larger timescale'' information, unless it is overdamped. Instead of the latter, it turns out that a \emph{drop clock} is a more natural model for this task.

More complicated models will exist in nature, but our question is what the simplest ones are. More complex clocks, such as clocks with time-dependent transition laws require more memory under the Markovian. Another example is that of gene regulatory networks which extends beyond discrete time into continuous time.

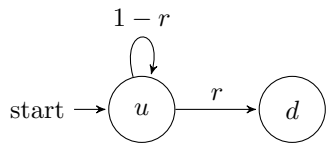
\begin{figure}[htbp]
    \centering
\begin{tikzpicture}[>=stealth',shorten >=1pt,auto,node distance=2cm]
  \node[initial,state] (u)      {$u$};
  \node[state]         (d) [right of=u]  {$d$};

  \path[->]
      (u) edge [loop above] node {$1 - r$} (u)
      (u) edge              node {$r$} (d);
\end{tikzpicture}
    \caption{The drop clock, a Markov chain with a probability $r$ to permanently transition to state $d$ and remain there.}
\end{figure}

The drop clock  starts (as all our clocks)  in a well defined state, namely $u$, i.e.\  $P(S=u | T = 0) = 1.$
After each time step, there is a probability  $r$ that the clock ``decays'' (transitions from state $u$ to state $d$) and a probability $1-r$ that it does not. Once the clock has decayed, it remains in state $d$ forever. The behaviour that this rule generates is an exponential decay $(1 - r)^t$ in time. An agent infers probable time only from the state of this clock.

\section{Experiments with 1-bit Clocks}
Our first experiments are dedicated to the 1-bit (i.e.\ 2-state) clocks. As discussed earlier, there are only few distinct classes of such clocks. Most notable are the 2-state alternator and the drop clock.
The alternator offers the maximum of 1~bit of information about time, namely whether one is in an odd or even time step. However, this information is purely local and cannot distinguish whether one is in an early or a later section of a run. However, even with only 1~bit of state, the drop clock can provide this distinction, albeit imperfectly at less than 1~bit resolution. For this to provide best results, the probabilistic decay rate of the drop clock (unlike the alternator) must be attuned to the length of the total time interval of interest; this rate will be in general acquired by evolution or some learning mechanism --- here, we will directly obtain it by optimization of informational costs. This is studied next. Our axis of time is almost featureless except for two features: length of time and the grain. The alternator matches the grain (local information) and the drop clock matches the length (global information). We do not impose any other features on the axis of time such as months or seasons because we only study pure time. For a meaningful link to external events, that would extend to the work of \cite{klyubin2007representations}.

\subsection{Measuring Large Time Scales}

\begin{figure}[htb]
\centering
\begin{tikzpicture}
\begin{axis}[
%      legend style = { at = {(0.95, 0.8)}},
%      %unit markings=slash space,
    xlabel=Timespan,
    ylabel=Optimal $r$,
    xmin=2,xmax=100,
    ymin=0,ymax=1.1,
    height=3.6cm,
    width=6cm
]

\addplot [] table[x index=0, y index=1] 
    {data/bestr_known_start_100_1000.dat};
% \addlegendentry{Drop clock}
% \addplot [dashdotted] table[x index=0, y index=1] 
%     {data/bestr_pre_decayed_100_1000.dat};
% \addlegendentry{Pre-Ticked Drop clock}
%    \addplot [dotted] table[x index=0, y index=1] 
%      {data/bestr_double_decay_100_10.dat};
%    \addlegendentry{Symmetrical Flip-Flop}
\end{axis}
\end{tikzpicture}
\caption{The optimal drop probability $r$ for different timescales. Note that the curve for the drop clock has a discontinuity at ${T \approx 15}$.}
\label{fig:drop-clock-phase-transition}
\end{figure}
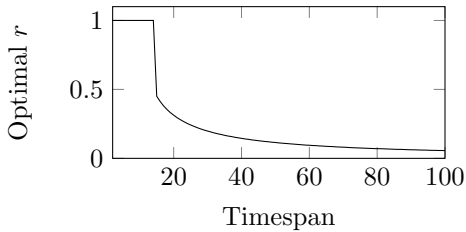

We now investigate time measurement by drop clocks at different timescales. We tune the clock by finding the drop probability $r$ maximizing $I(S; T)$ for the particular timescale of the experiment. Optimizing this by grid search gives Fig.~\ref{fig:drop-clock-phase-transition}.

\begin{figure}[htb]
\centering
\begin{tikzpicture}
\begin{axis}[axis on top,
  xmin=2,xmax=30,ymin=0,ymax=1,
  colorbar,
  colormap/bluered,
  xlabel=Timespan Size,
  ylabel=Drop Probability,
  x=5cm/30,y=4cm
]
\addplot graphics [xmin=2,xmax=30,ymin=0,ymax=1]
  {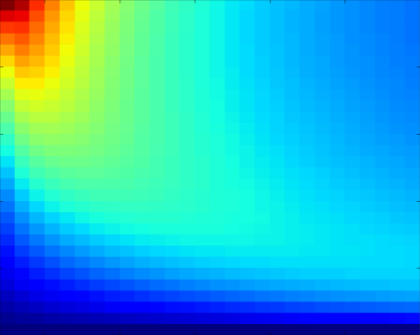};
\end{axis}
\end{tikzpicture}
\caption{Time information for different drop probabilities and time-spans.}
\label{fig:drop-clock-2d-plot}
\end{figure}

The first result is that the decay rate $r$ that best resolves time with a drop clock clearly depends on the time interval. Such a clock, therefore, must be adapted to the particular time interval to resolve. Strikingly, we find two regimes of solutions, namely one with one fixed decay rate (up to $T \approx 15$), and then a time interval-dependent decay rate.
An inspection of Fig.~\ref{fig:drop-clock-2d-plot} shows a relatively complex landscape where a global maximum of time information at the maximal decay rate $r=1$ is superseded at larger times by maxima at lower decay rates.

\begin{figure}[htb]
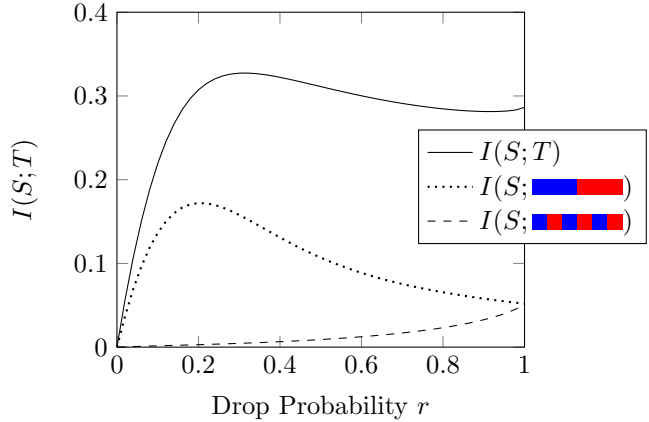

\centering
\begin{tikzpicture}
\begin{axis}[
  width=7cm,
  legend style = { at = {(1.3, 0.65)}},
  legend cell align={left},
  xlabel=Drop Probability $r$,
  ylabel=$I(S;T)$,
  xmin=0,xmax=1,
  ymin=0,ymax=0.4,
]

\addplot [] table[x index=0, y index=1] {data/known_start_reldrop_20.dat};
\addlegendentry{$I(S; T)$};
\addplot [dotted, thick] table[x index=0, y index=2] {data/known_start_reldrop_20.dat};
\addlegendentry{$I(S; \input{figs/lr-variable.tex}\unskip)$};
\addplot [dashed] table[x index=0, y index=3] {data/known_start_reldrop_20.dat};
\addlegendentry{$I(S; \input{figs/alt-variable.tex}\unskip)$};

\end{axis}
\end{tikzpicture}
\caption{Time information for different drop probabilities at ${T = 20}$. Apart from total time information, it shows information of clock state $S$ about being in an odd vs.\ even time step or  early vs.\ late phase of the measured period.}
\label{fig:two-maxima-drop-clock}
\end{figure}

Figure~\ref{fig:two-maxima-drop-clock} shows with a solid line a vertical slice from Figure~\ref{fig:drop-clock-2d-plot} at ${T = 20}$. \diffnew{This information curve does not have a unique maximum, but an inflection. We provide an explanation for the two local maxima by partitioning time into different features: one-way to look at time is to see an ``earlier'' and a ``later'' part and the other way is to look at ``odd'' vs ``even'' times. Our interpretation is that the different maxima come from information the clock has about different partitionings of time. The maximum around $r \approx 0.3$ appears to come from the global picture the clock has (the ``earlier'' vs ``later'' partitioning \crule[blue]{0.2cm}{0.2cm}\crule[blue]{0.2cm}{0.2cm}\crule[blue]{0.2cm}{0.2cm}\crule[red]{0.2cm}{0.2cm}\crule[red]{0.2cm}{0.2cm}\crule[red]{0.2cm}{0.2cm}\unskip) while the second maximum at $r = 1$ comes from the ``odd'' vs ``even'' partitioning (\crule[blue]{0.2cm}{0.2cm}\crule[red]{0.2cm}{0.2cm}\crule[blue]{0.2cm}{0.2cm}\crule[red]{0.2cm}{0.2cm}\crule[blue]{0.2cm}{0.2cm}\crule[red]{0.2cm}{0.2cm}\unskip).}

 Lower decay rates prove better at resolving global timing (early or late), and, while the drop clock is generally weak at resolving odd/even times, it still performs this resolution best for a hard decay of $r$, right at the beginning of the interval. This explains the phase transition in Fig.~\ref{fig:drop-clock-phase-transition} between the regime of small time-spans to that of large time-spans. This transition occurs because of the inflection in the information curve of the sharply initialized drop clock, when one maximum dips down below the other: different parts of the curve derive from knowing different things and because in some regimes one dominates the other.

We note that these different regimes make the drop clock hard to optimize --- small time spans require a different strategy than the long ones: not only must the clock be attuned to a particular time scale to best measure global time, but also does the information curve of the clock show two maxima where one overtakes the other. 
\subsection{Bag of Clocks}
Having studied the 2-state-clocks we now design a larger experiment. We would like to keep the clocks simple, but be able to measure time more accurately. For this purpose, we consider ``bags'' of independent clocks to measure time:
none of the clocks in the bag can communicate with any other clock, but to determine time, we consider the state of the whole clock collective.

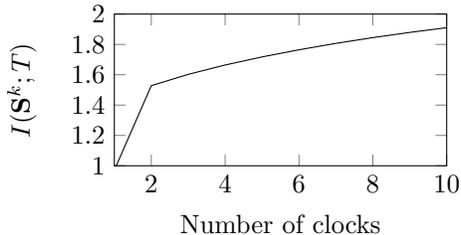
\begin{figure}[htb]
\centering
\begin{tikzpicture}
\begin{axis}[
  width=6cm,
  height=3.6cm,
  xlabel=Number of clocks,
  ylabel=$I(\mathbf{S}^k;T)$,
  xmin=1,xmax=10,
   ymin=1,ymax=2,
]

  \addplot [] table[x index=0, y index=1] 
    {data/darwin_mis_5.dat};
%    \addplot [] table[x index=0, y index=1] {data/parralel_10_60.dat};
%    \addlegendentry{$t_{max} = 5$};

\end{axis}
\end{tikzpicture}
\caption{Amount of information as the size of the collection grows for a time interval of length $T_{\text{max}} = 5$.}
\label{fig:frozen-bag-info-curve}
\end{figure}

The experiment is started without any clocks and the collection is built up incrementally, one clock at a time by optimizing the current clock under consideration. We use this incremental process to capture our intuition: that in evolution, existing features tend to be ``frozen'' because they are intertwined with the rest of the organism and new features tend to be optimized in relation to the existing frozen features. More precisely, assume $n$ clocks are already in the collection ($n=0,1,2\dots$). Given the state of the clock collection $\mathbf{S}^n = (S_1,\dots,S_n)$ with $n=0$  indicating the empty collection, the dynamics of clock $n$ is optimized with grid search as to maximize $I(\mathbf{S}^n;T)$, always for fixed total duration $T_{\text{max}}$; the dynamic parameters of all clocks $k=1,\dots,n-1$ are kept fixed during the optimization. Once the optimization is complete, a new clock $n+1$ is added and the procedure repeated. 
In our experiments, we stopped the process, once we reached 10 clocks. Importantly, in this experiment, the optimizer is allowed to choose any parameters of the dynamics, and the clocks that  the optimization finds are a pure alternator or else, drop clocks. As the collection grows, so does the achieved time information ${I(\mathbf{S}^n;T)}$ (Fig.~\ref{fig:frozen-bag-info-curve}). The first clock, with no oscillatory component added to the bag, is the alternator, as intuitively expected, as it resolves 1 full bit of information; but, notably,  all subsequent additions to the bag turn out to be pure drop clocks. The first two clocks add ${\sim 1.5}$ bits, while every subsequent clock adds significantly less.

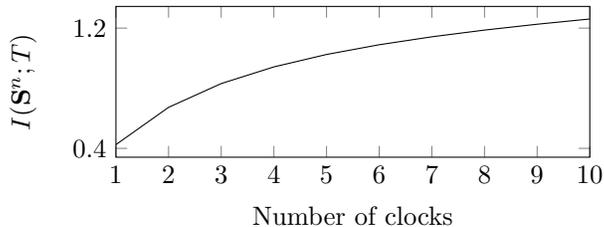
\begin{figure}[htb]
\centering
\begin{tikzpicture}
\begin{axis}[
xlabel=Number of clocks,
ylabel=$I(\mathbf{S}^n;T)$,
xmin=1,xmax=10,
y=2cm, ytick={0.4,1.2},
x=0.7cm, xtick={1, ...,10}
]

\addplot [] table[x index=0, y index=1] 
    {data/parralel_10_5.dat};
%    \addlegendentry{$t_{max} = 60$};
%    \addplot [] table[x index=0, y index=1] {data/parralel_10_60.dat};
%    \addlegendentry{$t_{max} = 5$};

\end{axis}
\end{tikzpicture}
\caption{Time information as more identical clocks are added to the set for $T_{\text{max}} = 50$.}
\label{fig:poisson-info-curve}
\end{figure}

In detail, we notice that all the clocks except for the first two have very similar dynamics (parameters not shown here). In other words, once the first two clocks are added, all further clocks essentially act together as an increasingly refined binomial process, as more clocks are added\footnote{Special thanks to Nicola Catenacci-Volpi for this observation.}.  Because the last clocks added to the bag have almost identical parameters, we ran another calculation to explore the behaviour of populations of identical drop clocks. We use a bag of pure drop clocks with the same $r = 0.1$, starting with one of these clocks and adding more until we reach a total of 10 clocks. We compute $I(\mathbf{S}^n;T)$ for a time scale of size $50$ to create Figure~\ref{fig:poisson-info-curve}. 

\subsection{Composite Clock}

Beyond the simplest clocks and unstructured clock bags listed above, we consider the next more complex clock, a composite clock consisting of two simple 1-bit clocks each, which, however, are permitted to communicate, but with a constraint on how much communication is permitted. This constraint is expressed as a measure of information flow between the composite  clocks.
This information flow constraint is how we penalize modular systems for their complexity. \diffnew{Consistent with the freezing procedure in the bag-of-clocks experiment we freeze the first clock here as well. Thus, the first (upper) component of the composite clock becomes an alternator.} Only the second (lower) component's dynamics will be parametrized and the parameters optimized according to suitable informational criteria. This one-way communication is inspired by the semigroup decomposition of counters \citep{rhodes2009applications}. Future work will allow the first clock to be optimized as well and add a feedback channel.

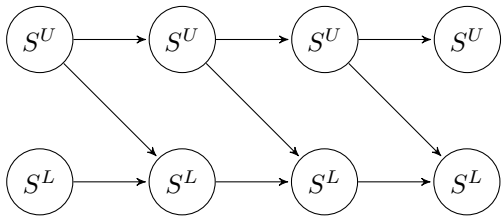
\begin{figure}[htb]
    \centering
\begin{tikzpicture}[
      >=stealth',
      shorten >=1pt,
      auto
    ]
    \node[state] (u0) {$S^U$};
    \node[state] (l0) [below=of u0] {$S^L$};
    \node[state] (u1) [right=of u0] {$S^U$};
    \node[state] (l1) [right=of l0] {$S^L$};
    \node[state] (u2) [right=of u1] {$S^U$};
    \node[state] (l2) [right=of l1] {$S^L$};
    \node[state] (u3) [right=of u2] {$S^U$};
    \node[state] (l3) [right=of l2] {$S^L$};

    \path[->] (u0) edge node {} (u1);
    \path[->] (u0) edge node {} (l1);
    \path[->] (l0) edge node {} (l1);
    \path[->] (u1) edge node {} (u2);
    \path[->] (u1) edge node {} (l2);
    \path[->] (l1) edge node {} (l2);
    \path[->] (u2) edge node {} (u3);
    \path[->] (u2) edge node {} (l3);
    \path[->] (l2) edge node {} (l3);
\end{tikzpicture}
    \caption{The structure of the composite clock unrolled in time. There is a hierarchy of information flow here. Both clocks send information to themselves but only the upper clock sends information to the lower clock.}

\end{figure}
To prepare the description of the full system, we write out in detail the first component of the clock as a Markov chain; it is  defined by the matrix $A_U = \left[\begin{smallmatrix}0 & 1 \\ 1 & 0\end{smallmatrix}\right]$ and initial state distribution\footnote{We use the random variable name as a proxy notation for the whole distribution.} $S^U_0 = \left[\begin{smallmatrix}1 \\ 0\end{smallmatrix}\right]$. In the next step, we design the dynamics of the lower component $S^L$, but it is not independent anymore. Rather, it can be influenced by the state of the upper component $U$. The matrix that will drive the behaviour of the lower component is a conditional probability distribution, a mapping from the joint state of both components $(S^U_t, S^L_t)$ at time $t$ to the future state of $S^L_{t+1}$ at time $t + 1$. There are 4 columns in this matrix because there are four combinations of the input (i.e.\ condition) states from both components: $uu$, $ud$, $du$, $dd$ (both up, one up and the other down, etc.).
In vector notation, the complete probability of the whole system to be in a state is written as:
\begin{equation} \label{eq:combine-states}
  P(S^U,S^L) = 
  \begin{bmatrix}
    P(S^U=u,S^L=u) \\
    P(S^U=u,S^L=d) \\
    P(S^U=d,S^L=u) \\
    P(S^U=d,S^L=d)
  \end{bmatrix},
\end{equation}
and the transition matrix for the lower clock (we remind that the upper clock is an alternator and that the first coordinate is the probability for $u$ and the second the probability for $d$) as $A_L = \left[\begin{smallmatrix}\theta_1 & \theta_2 & \theta_3 & \theta_4 \\ 1 - \theta_1 & 1 - \theta_2 & 1 - \theta_3 & 1 - \theta_4\end{smallmatrix}\right]$
and where we make use of the fact that the resulting probabilities add up to 1 (and we thus need only one parameter to describe each of the conditionals).
Combine now the matrix for the upper component ($A_U$) and the matrix for the lower component ($A_L$) to obtain the Markov matrix for the whole clock:

\[ 
A = 
 \left[\begin{matrix}0 & 0 & \theta_{3} & \theta_{4}\\0 & 0 & 1 - \theta_{3} & 1 - \theta_{4}\\\theta_{1} & \theta_{2} & 0 & 0\\ 1 - \theta_{1} & 1 - \theta_{2} & 0 & 0\end{matrix}\right].
\]

We  initialize the lower clock $S^L$, as always, in state $u$, i.e.\ with probability $ \left[\begin{smallmatrix}1 \\ 0\end{smallmatrix}\right]$. Now we have all required definitions. Expressed in terms of the joint variable $S= (S^U,S^L)$, and following the conventions from (\ref{eq:combine-states}),  the initial state of the complete clock is $S_0 = \left[\begin{smallmatrix}1 & 0& 0& 0\end{smallmatrix}\right]'$ (where $'$ is transpose). Using $A$, the matrix that ticks the clock forward to its next time step, we can simulate the clock starting at time 0 and  into a  future time $t$ by repeated matrix multiplication: $A^t$. For illustration,  the first few states look like:

\begin{figure}[htb]
    \centering
\def\arraystretch{1.2}
\begin{tabular}{ |p{0.6cm}|p{1.3cm}|p{5cm}|  }
 \hline
 \multicolumn{3}{|c|}{Time dynamics of both parts} \\
 \hline
 $S_0$ & $A S_0$ & $A^2 S_0$ \\
 \hline
 
 &&\\
 $\left[\begin{matrix}1\\0\\0\\0\end{matrix}\right]$ &
 $ \left[\begin{matrix}0\\0\\\theta_{1}\\1 - \theta_{1}\end{matrix}\right]$ &
 $ \left[\begin{matrix}\theta_{1} \theta_{3} + \theta_{4} \left(1 - \theta_{1}\right)\\\theta_{1} \left(1 - \theta_{3}\right) + \left(1 - \theta_{1}\right) \left(1 - \theta_{4}\right)\\0\\0\end{matrix}\right]$ \\
 &&\\

 \hline
\end{tabular}

    \caption{The probabilistic state of the composite clock for $t = 0$, $t = 1$ and $t = 2$.}
\end{figure}

\subsection{Experiments with the Composite Clock}
The current experiments distinguish themselves from the earlier ones by having the participating clocks communicate: we create an information ``tap'' between the two clocks. Importantly, we will  limit the amount of information that may flow from the upper to the lower clock. As discussed earlier, all costs/rewards (e.g.\ flow vs.\ time resolution) will be expressed exclusively in terms of Shannon information as unique currency of discourse.

In the same spirit as in the bag-of-clocks experiment,  we greedily pre-optimize the upper clock, which results in an alternator whose parameters will be frozen. Thus, the experiment will only optimize the remaining parameters of the clock, i.e.\ the dynamics of the lower clock and the parameters of its dependence on the upper clock. 
While we maximize time information as before, the constraint $C$ is put on the capacity of communication (\emph{transfer entropy}) from the upper clock to the lower one ($T$ in the index denotes  marginalization over all valid times occurring in the experiment). Since the system is Markovian, flow only over a single step is considered. Since the upper clock is fixed as the alternator, no feedback channel is included. We compute $\max_{I(S^U_T;S^L_{T+1}|S^L_T) \leq C} I(S;T)$.

\begin{figure}[htb]
\centering
\begin{tikzpicture}
\begin{axis}[
    xlabel=Information Flow Constraint $C$ (in bits),
    ylabel=$I(S;T)$,
    xmin=0, xmax=1,
    ymin=1.4,ymax=2
]

%% \addplot [color=red] table[x index=0, y index=1] 
%%     {data/drag_n9_mi_condmi_a.dat};
\addplot [] table[x index=0, y index=1] 
    {data/drag_n9_mi_condmi_a1.dat};
\addplot [color=red] table[x index=0, y index=1] 
    {data/drag_n9_mi_condmi_a2.dat};
\addplot [] table[x index=0, y index=1] 
    {data/drag_n9_mi_condmi_b.dat};
\addplot [] table[x index=0, y index=1] 
    {data/drag_n9_mi_condmi_c.dat};
\addplot [mark=*, color=red, mark size=1.5] table []
{
0.9999999999999487   1.9749375012017316
};

\addplot[mark=*, color=red, mark size=1.5] coordinates{(0.0421478658,1.5063372572)};
\addplot[mark=*, color=red, mark size=1.5] coordinates{(0.2400902399,1.7425492069)};

\end{axis}
\end{tikzpicture}
\caption{In red, the optimum curve found by the DIRECT global Lipschitz optimizer. In black, the more complete curve found by the COBYLA local optimizer.}
\label{fig:time-information-coupled-clocks}
\end{figure}
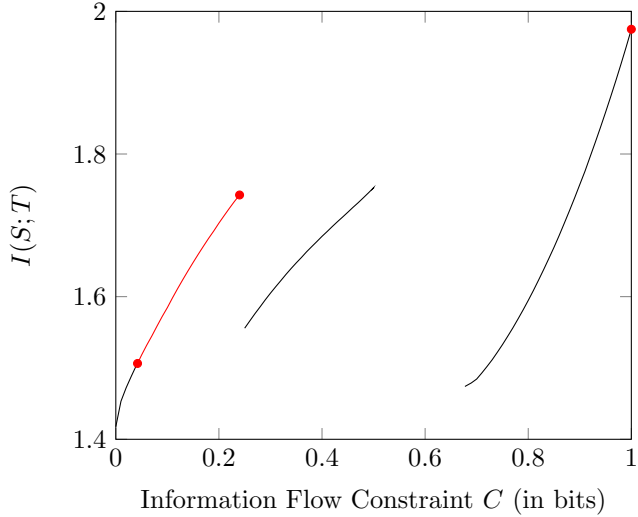

To optimize with the constraint, we use the Lagrangian method,  i.e.\ we optimize $I(S;T) - \lambda I(S^U_T;S^L_{T+1}|S^L_T)$.  Scanning through the possible values of $\lambda$, we cover the spectrum of clocks arising through possible constraints. We first used the DIRECT Lipschitzian method for global optimization \citep[][shown in red in Fig.~\ref{fig:time-information-coupled-clocks}]{jones93:_lipsc_optim_lipsc_const}. One finds two distinct regimes: the perfect clock at $C=1$, and distinctly suboptimal clocks in the regime below around $C\approx0.2$. The curve breaks off at the low end at about $C=0.04$ due to memory limitations of the Lipschitzian optimizer.

The other optimizer used was COBYLA \citep{powell94}, where we used a relaxation method to cover the curve. We started optimization at the permissive information flow bound $C=1$, optimized, then the bound was slowly tightened, always starting with the clock parameters obtained for the previous constraint $C$. The results are shown again in Fig.~\ref{fig:time-information-coupled-clocks}, but include the black regions in addition to the red. This optimization being not global, it uncovers additional structure in the solution space. First of all, when, starting at the optimal clock $C=1$, $C$ is reduced, the trade-off curve between the constraint $C$ and the time information $I(S;T)$ falls below the convex hull of the (globally optimal) red solutions. These black points will thus never be found by a global optimization of the Lagrangian. They correspond to ``fuzzy'' counters (the optimal clock is a binary counter), but do not trade in information flow and achieved time information in a well-balanced way, although part of that portion of the curve is still Pareto-optimal (is not superseded simultaneously by solutions better in terms of both $C$ \emph{and} $I(S;T)$). 

In the range $C\approx0.5-0.7$, no solution is found. Between $C\approx0.25-0.5$, a class of locally optimal solutions is found which is neither findable by the Lagrangian, thus not on the convex hull, nor is it Pareto optimal. Finally, below $C=0.25$, one regains the lower $C$ regime from the global optimization. The curve continues down to $C=0$ (continuation of the red to the black curve), not suffering from the memory problems of the global optimizer. The two clock classes below $C\approx0.5$ look very similar as a whole clock, but distribute the ``counting'' differently over their component clocks.

Apart from the discovery of different clock ``regimes'', detailed inspection of the space of possible configurations (not shown here) demonstrate that, in part, finding parameters which achieve high values of time information  is difficult and, despite large permitted flows $C$, time resolution may still have low values. This makes it clear that measuring time is not an incidental effect that is likely to  be found en passant, but one rather will expect good time measurement abilities to have been  optimized for, either directly or indirectly (via proxy criteria). In any case, the clocks in the hierarchy need to be attuned to each other for optimal effect.

\section{Final Comments and Future Work}

We have studied the ability of minimal clocks to resolve time information. In particular, even 1-bit clocks can, as drop clocks, provide information about global time if the overall time horizon is known when the clock parameters are set. The relation of measured interval and clock parameters is not straightforward and marked by an interplay of global and local properties (Fig.~\ref{fig:two-maxima-drop-clock}). When extending to a bag-of-clocks, the first two clocks are an alternator and a drop clock, but all further drop clocks are nearly identical, so that apart from the first two, the rest of the clocks operate as a nearly binomial process. Finally, when two clocks are stacked together with limited communication, a rich set of regimes opens up, of which just two, the perfect clock, and a very soft clock, can be Lagrange-optimal. These studies provide a spectrum of candidates for behaviours of minimal clocks which one could try to identify in biological systems. 

One limitation of the present work is the assumption of a fixed, global tick which drives all the clocks. Future work will, therefore, include the consideration of clocks in continuous time, where the dynamics needs to establish and sustain a synchronization between the subclocks in addition to the coordination of their respective resolution regime. 

We conjecture that informationally optimal clocks will exhibit some robustness. Applying a criterium of robustness to them is left for future work.

\section{Acknowledgements}
We thank Nicola Catenacci-Volpi, Simon Smith and Adeline Chanseau for discussions on this topic.

Christoph Salge is funded by the EU Horizon 2020 programm under the Marie Sklodowska-Curie grant 705643.

\footnotesize
\bibliographystyle{apalike}
\bibliography{references}

\end{document}